\documentclass[]{aastex} \usepackage{amsmath}

\def\deg{^\circ\!}
\def\bef{\begin{figure}}
\def\eef{\end{figure}}
\def\beq{\begin{equation}}
\def\eeq{\end{equation}}
\def\ber{\begin{eqnarray}}
\def\eer{\end{eqnarray}}
\def\ben{\begin{enumerate}}
\def\een{\end{enumerate}}

\def\beb{\begin{thebibliography}{99}}
\def\eeb{\end{thebibliography}}

\def\etal{{\it et al.\/$\!$}}
\def\eg{{\it e.g. }}

\def\viz{{\it viz. }}

\def\la{\mathrel{\hbox{\rlap{\hbox{\lower4pt\hbox{$\sim$}}}\hbox{$<$}}}}
\def\ga{\mathrel{\hbox{\rlap{\hbox{\lower4pt\hbox{$\sim$}}}\hbox{$>$}}}}

\def\mushroom{\mbox{GW123.4-1.5}}

\def\ir10025{$I_{100}/I_{25}$}
\def\IRR10060{$I_{100}/I_{60}$}
\def\IR2512{$I_{25}/I_{12}$}
\def\60100{$I_{60}/I_{100}$}
\def\1225{$I_{12}/I_{25}$}
\def\kmss{$\rm km \ s^{-1}\ $}
\def\kmsp{$\rm km \ s^{-1}$}

\slugcomment{To be submitted to AJ}

\shorttitle{search for vertical structures... CGPS data}
\shortauthors{Asgekar et al.}

\begin{document}


\title{A Search for Narrow Vertical Structures in\\
the Canadian Galactic Plane Survey}


\author{Ashish Asgekar\altaffilmark{1}, 
Jayanne English, Samar
Safi-Harb\altaffilmark{2} }
\affil{Department of Physics and Astronomy, University of Manitoba,
    Winnipeg, MB R3T 2N2, Canada}
\and 
\author {Roland Kothes}
\affil{National Research Council of Canada, Herzberg Institute of Astrophysics, 
Dominion Radio Astrophysical Observatory, PO Box 248, Penticton, British Columbia, V2A 6J9, Canada and the Department of Physics and Astronomy, University of Calgary, 2500 University Drive NW, Calgary, AB, T2N 1N4, Canada}
\email{ashish@physics.umanitoba.ca, jayanne\_english@umanitoba.ca,
safiharb@cc.umanitoba.ca, Roland.Kothes@nrc-cnrc.gc.ca}


\altaffiltext{1}{Currently at Raman Research Institute, Bangalore, India 560 080.}
\altaffiltext{2}{NSERC/University Faculty Award fellow}


\begin{abstract}

Worms are defined to be dusty, atomic hydrogen (HI) structures which
are observed in
low resolution data to rise perpendicular to
the Galactic plane. Data from the $1^{\prime}$ resolution Canadian
Galactic Plane Survey (CGPS) were systematically searched for narrow
vertical HI structures which could be resolved worms.  Another motivation
for the search was to explore the scenario that mushroom-shaped worms 
like \mushroom, studied by English and collaborators, 
could be generated by a single supernova.  However no other vertical
structures of mushroom-shape morphology were found. We also examined
objects previously classified as worm candidates by Koo and
collaborators; only 7 have a significant portion of their structure
falling in the CGPS range of $l=74\deg$~to $147\deg$, $-3.5\deg < b <
+5.5\deg$.  Apart from \mushroom\ we could not confirm that any of
these are coherent structures that extend towards the Milky Way's
halo.  However a list of 10~narrow, vertical structures found in our
search is furnished; one structure is~$\gtrsim500\,$pc tall, thus
extending from the Galactic plane into the halo. We provide details
about these narrow vertical structures, including comparisons between
HI, radio continuum, IR, and CO observations.  Our search was
conducted by visual inspection and we describe the limitations of this
approach since it indicates that only 6~disk-halo features may exist
throughout the Milky Way. We also discuss possible origins of
structures at high latitudes and the relationship between mushroom-shaped
clouds and old supernova remnants. 

\end{abstract}




\section{Introduction}
\label{s:intro}

Atomic hydrogen (HI) gas structures called 
shells, chimneys, and worms are superimposed on 
our Milky Way's interstellar
gas disk, which in turn displays an enhanced spiral structure and
has a Gaussian Full-Width-Half-Maximum (FWHM) thickness of $\sim 210$\,pc
\citep{dic90}. Some of these structures
could be relics of violent phenomena such as stellar winds
and supernov\ae~(SNe) which continually churn up and
inject energy into the interstellar medium (ISM). For
example,  young star clusters are expected to excavate
large cavities, or superbubbles, in the Galactic plane.
Chimneys are theorized to be an evolutionary
phase in the lifetime of these superbubbles and to act as
conduits which transport ionizing radiation and hot
gas to the halo of the Milky Way. The presence and
maintenance of warm and hot ionized gas far above
the plane arises naturally in this ``chimney model"
of the interstellar medium \citep{nor89}.

More specifically, chimneys are expected to evolve out of supershells,
which are HI  skins of superbubbles consisting of
hot gas generated by young stars and SNe. Often, supershells
are identified as voids in the spectral-line maps of HI and have sizes of
$\sim 500$\,pc \citep{mcg00}. These structures are too big to be
attributed to a single supernova explosion or winds from a single
star. However, simple scenarios which use multiple SNe
or clustering of O-type stars  in order to form 
thin-shelled superbubbles \citep{mac89}, still can not account for the
sizes and energy requirements of some observed supershells
\citep[\eg$\!\!$][]{hei84}.

\citet{hei84} and later \citet[][hereafter KHR]{koo92}
used spatial filtering of velocity-integrated HI line
data, along with its correlation to infrared (IR) data,
to discover dusty, wiggly gas elements ``crawling away
from the Galactic plane". These were termed ``worms"
and their origin has not been fully explained.
At least some were believed to be post-cursors of
broken walls of HI supershells. Thus, worms could be
conduits in transferring ionized matter from the disk
to higher Galactic latitudes, and the warm ionized
medium \citep{rey89,rey91} was proposed to be related
to the observed ionized gas associated with these worm
candidates \citep{hei96}. Based on the distribution of
candidates in their survey, KHR estimated a total of
50-100~chimneys in the Galaxy. However, the KHR catalog
may be affected by the confusion of intrinsically
separate HI clouds. For example, approaches similar
to KHR's methodology that have been applied to other
data sets, \eg~the Leiden-Dwingloo Survey over
the CGPS range \citep[LDS,][]{bur94}, have not
recovered most of
the worms cataloged by KHR and have generated other
worm candidates not found previously \citep[][]{nor03}.

A KHR worm candidate, GW123.4-1.5, was studied using
HI line data collected from 0$\deg$~to $6\deg$~below the Galactic
plane with the synthesis telescope at Dominion Radio
Astrophysical Observatory \citep[DRAO, ][]{lan00}.
These data were contiguous with the Canadian Galactic
Plane Survey \cite[CGPS,][]{tay03}, which terminated
at~$b\sim-3.5\deg$. The study revealed that it resembled
a mushroom in its appearance, with a stem and a cap
\citep{eng00}. Although superbubbles created by winds
from multiple stars could appear  mushroom-like
\citep{ten87, mac89}, this worm is not consistent
with these classical theories, since more mass is in
the cap than in the stem~(3:1). Additionally, the
classical models only form caps in the halo while the
top of the structure is 420~pc away from the Galactic
plane. This height is about 80~pc short of the 
canonical 500-pc
altitude that is 
associated with the commencement of the
halo in HI observations \citep{loc86}. The
kinetic energy of the associated HI gas is estimated
to be $1\times10^{50}$~ergs,
which indicates that only an energy input equivalent
to that of a single supernova is necessary to result
in such an object. This modest energy budget allows
for alternative scenarios. For example, the ``mushroom"
could be produced by the impact of an infalling cloud
\citep{tak03} or the buoyant rise of hot gas, including
gas accumulated in the Galactic plane \citep{dea01}.
Also, rather than being in the class of large cluster-blown
supershells (such as GW162.6+4.9, another KHR candidate
of length $\sim 1$\,kpc), GW123.4-1.5 may have evolved
from their smaller cousin, a single supernova event.
This further suggests a reinvestigation of the origin
of worms using high-resolution CGPS data. 

The mushroom-shaped cloud may still be able to
transport matter into the halo. Two numerical models
\citep{eng00,dea01} associate the cloud with a bubble
of hot gas rising away from the disk buoyantly.
\citet{eng00} modeled the evolution of a single,
off-plane supernova. In the scenario by \citet{dea01},
hot gas from dispersed SNe and hot young stars collects
in a reservoir and subsequently a bubble of ionized gas
rises from this pool buoyantly. Both scenarios indicate
that a buoyant structure evolves into a mushroom shape
in 3-8~Myr. 

We would like to test the single supernova scenario above.  We suspect
that some correlation would then exist between the properties of
GW123.4-1.5-type worms and late-type supernova remnant (SNR)
shells. In particular, we ought to observe a transition phase between
the morphology of late type SNRs \citep[\eg~G55.0+0.3, ][]{mat98},
with ages around 2--7~Myr, and HI clouds that are just developing a
mushroom shape. While SNRs of ages greater than 4~Myr are only
proposed in theory, expanding HI shells, possibly due to SNRs of age
3--4\,Myr, have been detected and characterized as a part of the
Phase-I data of the CGPS \citep[CGPS-I, ][]{nor00, sti01}.  The
expanding shell GSH138--01--94, with a radius of about 180~pc, has
been interpreted as bubble powered by the joint action of strong
stellar winds from early-type stars and SN explosions \citep{uya02} or
as an old SNR expanding in a low-density, low-metallicity environment
\citep{sti01}.  The latter view indicates that comparisons between
SNRs and worms within a homogeneous data set, though difficult, are
possible.

As an initial step we have searched for narrow vertical structures and
mushroom-shaped structures in the CGPS-I data.
We found 10 vertical structures; these particular objects
did not allow us to characterize them according to their
plausible evolutionary sequence towards acquiring their mushroom
shape, and subsequent correlation with late stages of SNR
shells. However, this paper catalogs these structures, examines the
issue of confusion between sources in HI data cubes, and considers why
vertical structures are rare in a data set with $1\arcmin$ resolution.

In this search, we systematically examined the HI cubes from the CGPS-I
data {\it visually} (\S~\ref{s:mom_maps}) with the aim of detecting
narrow HI features that were more or less perpendicular to
the Galactic plane and which displayed coherence in velocity.
The attention to
velocity signatures was motivated by the velocity gradient displayed
by \mushroom\ (\S~\ref{s:template}).  In \S~\ref{s:template} we also
describe our examination of the KHR worm candidates. Next we describe
our approach for examining the local standard of rest (LSR) velocity ranges
in which a vertical structure with a height of 500~pc should be
resolvable~(\S~\ref{s:refined-search}), detecting no new
candidates. We then outline our method (\S~\ref{s:measure}) in
deriving the properties of our 10~candidate structures.

Our results~(\S~\ref{s:results}) include comments on individual
objects, including comparisons with IR, radio continuum and CO
emission and our assessment that none of the KHR worm candidates
falling within the CGPS range, other than \mushroom, are coherent
structures. We include in our list of detections 8 non-vertical
interesting HI features we encountered in the data. These, though
not relevant to our study, may be important in their own right
and of interest to some readers. In \S~\ref{s:discuss},
we estimate the number of narrow, vertical structures in the
Milky Way and the challenges associated with detecting these.
We also
review possible formation scenarios of structures extending into the
halo, and briefly outline future work.

\section{Data}
\label{sect:search}

\subsection{CGPS Data Sets}
\label{s:CGPS}

The CGPS is a project that coordinates the collection of
radio, millimeter and IR data surveying
73 degrees of  the Galactic
plane at arc-minute spatial resolution~\citep{tay03}. 
DRAO component of CGPS data consists of synthesis HI
spectral-line observations; radio continuum images at
1420~MHz including linear polarization; and total power
radio continuum images at 408~MHz. Data from the 26-meter
dish of the DRAO, which provided zero-spacings for the
HI line data, were also released as an
independent survey \citep[hereafter LRDS,][]{hig00}.
The CGPS collection includes the following data sets processed to match the 1
arc-minute resolution at 21-cm: molecular-line ($^{12}$CO) data from
the Outer Galaxy Survey (OGS) by the Five College Radio Astronomy
Observatory \citep[FCRAO,][]{hey98}; IRAS Galaxy Atlas \citep[IGA,][60
\& 100 $\mu$m bands]{cao97}; and the Mid-Infrared Galaxy Atlas
\citep[MIGA,][12 and 25~$\mu$m]{ker00}.

Briefly, the DRAO 21-cm data sets cover the Galactic
mid-plane from $-3.6\deg < b < +5.6\deg$ (Galactic
latitude) and $l=74.2\deg$~to $147.3\deg$ (Galactic
longitude); the HI spectral-line data cubes cover
LSR velocities $-150 < v < 50$~\kmss with a velocity
resolution of $1.3$~\kmss (channel width of~$0.82446$~\kmss).
The 74~cm data have a $\sim3.4 \arcmin$  resolution
and  span $-6.7\deg < b < +8.7\deg$. A single-channel HI
line map, for a declination $\delta$,  has a rms level in
brightness temperature of $3.5\,sin\delta\,$K 
whereas the typical rms levels in radio continuum maps are 
$71\,sin\delta\,$mK and  $750\,sin\delta\,$mK at 20~cm and
74~cm, respectively. In all these maps the rms varies by
$\sim 1.6$~over the CGPS mosaic area from its lowest value
at the center of each synthesis field. The FCRAO data only
cover $l=102.5\deg$ to $l=141.5\deg$  and $b=-3.03\deg$ to
$b=+5.6\deg$ and v$=-150\,$\kmss to +40\kmsp. We used the
version of the IRAS data that had been reprocessed to
1~arc-minute resolution for the entire CGPS spatial range.

For our initial search and analysis we used the CGPS HI line data
mosaics, which consist of 2.5$\deg$ diameter synthesis fields combined
into 5.12$\deg \ \times$ 5.12$\deg$ areas. Subsequently we searched
the other CGPS data sets for associated emission in radio continuum,
IR, and CO.  Where fields larger than the CGPS mosaics were required,
supermosaics were generated using DRAO's software. We
also searched the LRDS for extended vertical structures.

We also refer to phase II of the CGPS, currently underway, which is an
extension of the DRAO survey in both Galactic longitude and latitude.
CGPS-II is one component of a campaign which also uses the VLA and the
Australia Telescope National Facility's Compact Array. The project as
a whole is called the International Galactic Plane Survey (IGPS).

\subsection{Other Data Sets}

We examined the Leiden-Dwingloo Survey \citep[][]{bur94} for our
comparison with previously studied worms.  For multi-wavelength
comparisons associated with features spanning the larger field of view
of this and the LRDS, we used data sets available at the SkyView
database\footnote{\url{http://skyview.gsfc.nasa.gov/}} \citep{mcg96}.
Additionally we used the IRAS fields accessible from SkyView for our
examination of apparently old SNRs outside the CGPS field of view.

\section{Analysis}
\label{s:analysis}
\subsection{The Search for Narrow, Vertical Structures}
\label{s:mom_maps}

Our general strategy for detecting features rising from the mid-plane
towards the halo relied on visualization tools and is outlined here.
Our motivations and various approaches are detailed in the following
subsections. Individual candidate narrow, vertical structures are
described in \S~\ref{s:results} along with the resulting measurements
and derived properties (in Tables~\ref{t:results} and~\ref{t:results2}).

We began by using the 
Karma\footnote{\url{http://www.atnf.csiro.au/computing/software/karma/}}
 software~\citep{goo96} in order 
to search for narrow, streaming structures. Those rising away from the
plane, more or less at a right~angle, will have a fountain-like
appearance when the channel maps are viewed in a movie-like
manner in animation tools. Although we used logarithmic stretches
for intensity displays, we did not integrate over channels or
smooth in the spatial dimensions.

Our impressions of a coherent structure could be confirmed
using integrated HI brightness temperature maps and
intensity-weighted mean velocity maps
\citep[zeroth and first moment, respectively;][]{vang89}.
(Note: Although we refer to them as moment~0 maps,
multiplication by the velocity width of a channel
was omitted by the moment function in tool `KPVSLICE'
at the time of our measurements. Thus our integrated
HI brightness temperature maps, in units of kelvin, simply
sum intensity over channels.) For example, we constructed
a number of integrated HI brightness temperature maps over
the entire velocity width of an apparent structure. Each
map covered $\lesssim 5\,$\kmss and did not overlap with
the velocity range of the adjacent maps.  Since a cohesive
structure would be present over a velocity range greater
than that of the turbulent velocity \citep[5-10~\kmsp,
e.g.~][]{bel84}, we expected it to be present in more
than one successive map. Additionally we expected it to appear
spatially contiguous (that is without significant gaps) in adjacent
channels over its velocity range, unless it had a low signal to noise
ratio.  We additionally constructed an intensity weighted mean
velocity map of the entire structure. We would expect a coherent
structure to have a systematic variation in velocity that could be
distinguished from small-scale turbulent motions.  For example, small,
unrelated clouds could spatially overlap throughout some channels and
thus appear as one single object in an integrated HI brightness
temperature map;
however these clouds are expected to produce a pseudo-random
appearance in the mean velocity map. A detailed example of how
moment maps are used is provided in \S~\ref{s:KHR_null} which describes
the results of our examination of 
KHR worm candidates.

Thus our various examinations of the CGPS HI data cubes included the
following steps:
\begin{enumerate}
\item Individual channel maps were viewed using
animation software in order
to pick out any narrow vertical structures with a length:width ratio
exceeding 3:1. No other morphological distinctions were considered. 
\item Objects were retained that were discernibly
`connected'
over a minimum velocity range of 10~\kmsp. That is, components of a single 
structure were expected to
be spatially related throughout adjacent channels. For example,
a particular shape that appeared to move sideways as we stepped through
the channels would be considered connected. 
\item Continuity
over the entire length of the structure was ensured using 
integrated HI brightness temperature maps and 
mean velocity maps, as described above. 
\item The structure was examined (using the moment maps and 
position-velocity diagrams) for interesting velocity signatures,
such as expansion in the direction
of its  width  or flow along its length.
\item Regardless of whether a velocity signature was apparent or not,
IR and radio continuum images and CO data were overlaid on HI
integrated HI brightness temperature maps in order to search
for associations between the neutral gas, ionized gas, and dust
features.
\end{enumerate}

\subsubsection{Searching the Full Spatial and Velocity Range of 
the Data}
\label{s:template}

As well as sifting through individual fields of the entire CGPS-I, we
applied the steps in the previous section to three approaches intended to
find structures of any height which extend vertically away from the
midplane. As described below, we searched for HI features with some
characteristics similar to \mushroom; we examined KHR worms in detail;
and we searched for mushroom-cap structures in IR data.

We initially focused
on searching for \mushroom-type vertical structures whose origins 
require only
a modest input energy (e.~g.~that of a single SN event). 
Hence the development of the 
strategy outlined in the previous section is influenced by
observations of \mushroom. This cloud has a narrow stem with a
width-to-length ratio less \mbox{than 1:5}. The stem `rises' $\sim
200$~pc roughly perpendicular to the Galactic plane, and it shows a
systematic velocity gradient along the length of $12\pm3$\,\kmsp.
The $\sim$40-pc width of the stem also has a velocity signature
such that emission
at the center is red-shifted compared to the edge, which indicates
either expansion or contraction. The cloud's cap is about
130~pc\,$\times$\,130~pc wide and 
displays little velocity gradient in latitude or
longitude. There is a column-density contrast
of~1:2 between the surrounding HI and \mushroom. 
All these characteristics, including these distinct
velocity signatures (illustrated in Figure~\ref{f:mush-vel-sig}),
formed a template that was used to search 
for similar structures in the data.

\notetoeditor{Please put Figure~\ref{f:mush-vel-sig} here.}

The examination of the KHR worms (see Table~\ref{t:KHR_null}) included
correlating IR maps with the integrated HI brightness temperature HI maps.
This was motivated by the fact that \mushroom, which was included in their
catalog, has significant IR emission associated with its cap region.
KHR characterized their candidates with \60100~ratio, however the cap
of \mushroom\  is dominated by emission from the foreground star,
$\gamma$~Cass, in the \60100-ratio map. This is avoided in the
\ir10025 map of \mushroom, where  the cap morphology appears
most distinct from the ambient emission in the ratio range~12--37 (see
Figure~\ref{f:mush_IR100-25}).  

\notetoeditor{Please put Figure~\ref{f:mush_IR100-25} here.}

Note that without making comparisons
with HI emission the \mushroom\ cap is difficult to distinguish from
the ambient IR emission using IR maps or IR ratio maps.
Nevertheless we additionally searched for `cap'-like structures in the
CGPS IR mosaics and the infrared-ratio maps, particularly using
$\!\!$ I$_{100}$/I$_{60}$ maps.

While we did compare CO data with detected HI narrow vertical
features, none of the CO clouds from the region of the OGS
(\S~\ref{s:CGPS}) that overlapped with the CGPS clearly correlate with
the stem of \mushroom.  Hence, we did not use evidence of a
correlation between CO and HI as a criterion for classifying CGPS HI
structures as potential disk-halo interaction objects.

\subsubsection{Searching the velocity range associated with
  objects 500~pc tall}
\label{s:refined-search}

The data were also examined for HI structures that would specifically 
act as conduits in the disk-halo interaction. Such objects would rise
from the Galactic plane up to a height of $\sim 500\,$pc, the
canonical height for the commencement of the halo with respect to the
HI disk. A feature of this extent will be entirely visible within
$ b > -3.5\deg $ if it is at a minimum distance of 8~kpc. If the
object is any closer  we
may only see a part of the structure.  If it is mushroom-shaped we may
not see the cap, and we may not identify a stem as different from the
turbulent HI clouds in the field.  
For example, one of the reasons
\mushroom\  was identified as a mushroom-shaped cloud, was the
acquisition of synthesis observations of the cap to $b < -5\deg$,
which is outside the range of the CGPS coverage of the plane.
Other examples of this problem are provided in \S~\ref{s:systematics}. 

Assuming a structure has similar dimensions to \mushroom, then due to
the finite resolution element ($1\arcmin$) of the CGPS, a stem 40-pc
wide will not be adequately resolved (6$\times$ resolution element)
beyond a distance of 23~kpc from the sun. Given that this is the extent of the
Milky Way's disk, from a distance of $\sim 8\,$kpc onwards we should
find mushroom-shaped candidates, under otherwise perfect
circumstances. 

Using the \citet{bra93} model of the rotation curve of the Milky Way,
we computed the values of $v_{LSR}$ at different longitudes
corresponding to the minimum distance that a mushroom-shaped candidate
will be apparent in its entirety.   Table~\ref{t:refined-search}
summarizes the range of distances and velocities which were visually
examined.

\subsection{Measurements and  Derived Quantities}
\label{s:measure}
Once a candidate was identified through inspection of the
HI cube, we measured the HI column density, the total amount of
HI, the velocity characteristics, and dimensions.
The measurement techniques are presented here and the values
are presented in Tables~\ref{t:results} and \ref{t:results2}.

The FWHM velocity range and velocity centroid were estimated using
multiple position-velocity plots
across and along the length of a structure.  If emission from
unrelated structures in adjacent channels hindered this approach, then
the FWHM was estimated visually by stepping through channels. An
independent check was performed by creating velocity profiles through
various points  in the structure. The
FWHM has a typical error of~$\sim 2\,$\kmsp.

Integrated brightness temperature maps were used to measure
the extent of an object and to produce a column density map.
We used the zeroth moment map with no clipping of intensities
per channel, in order to sum the intensity values over the
entire velocity range determined by visual inspection. The
spatial dimensions were measured from this integrated HI
brightness temperature map using intensity profiles along
and across the structure; the structure's boundaries were
defined as having intensity values of 5 $\sigma$
above the background in  the surrounding regions.

We converted an integrated HI brightness temperature map
(units of~K) to a column density map (units of atoms~cm$^{-2}$)
in the usual way for optically thin HI\footnote{All
arithmetic operations were performed using Astronomical
Information Processing Software, AIPS++
(\url{http://aips2.nrao.edu/docs/aips++.html} .)}
\citep[see eq.~7.2][]{but88}. We enclosed
the structure as tightly as possible with a polygon and
computed the total number of pixels and the sum (total
integrated column density in the area), mean (column density),
and rms of the column density values. The selection of the
polygonal area is somewhat arbitrary and leads to the
noticeable error described below in~\S~\ref{s:errors}.

For the column density in Table~\ref{t:results2} we
have not removed a ``background component" representing
the ambient emission. The spatial area surrounding an
object within a channel, as well as channels adjacent
to the FWHM range, contain much structure making the
assessment of the ambient threshold difficult. We
include this omission in our uncertainty, see~\S~\ref{s:errors}.

We wrote a computer program based on the Galactic rotation model of
\citet{bra93} in order to estimate the distance of an HI object using
its LSR velocity and its Galactic longitude. (See \S~\ref{s:errors}
for associated errors.)  For each object we used this derived distance
to convert from the angular scale to the spatial scale of the
object.

In order to compute HI masses we use the sum of values within the
polygon. Since this measures the total column density in the structure
integrated over the velocity width, we only multiplied by the square
of the distance and by the spatial area of a pixel
(18$^{\prime\prime}$ $\times$ 18$^{\prime\prime}$ converted to
cm$^2$). This generated the total number of atoms which was
subsequently multiplied by the mass of a hydrogen atom and converted
to solar masses.  This is the first HI mass listed in
Table~\ref{t:results2} and is an overestimate since we had not removed
a contribution from the ambient medium.

An estimate of background contamination is unreliable since the
objects of study are usually crowded by extensive non-vertical
structures of similar HI column density. In spite of the difficulty in
selecting a portion of a field free of such structure, we quote in
Table~\ref{t:results2} a ``revised'' HI mass in which the contribution
from the ambient medium has been estimated and subtracted.  That is,
for each object we located, as best we could, a part of sky devoid of
large structures in the vertical feature's column density map, and
estimated the average column density over that area. Assuming that
this background is uniform across the structure of interest, we
multiplied the background average column density with the spatial area of the
vertical structure to obtain the total background column density.
This background column density was converted into HI mass and 
subtracted from the original HI mass estimate. 

\subsubsection{Errors in derived properties}
\label{s:errors}
 
In our distance estimation computer program the distance 
associated with each input
longitude is varied in linear steps. For each step we
compute the Galactocentric rotation speed \citep[given by][]{bra93}
and compute the LSR velocity ($v_{step}$).  If
$v_{step}$ matches the input velocity to within 0.1~\kmsp, we assign
the respective distance~($d$) to that velocity. This algorithm has a
typical error of 0.1~kpc, without any systematic variation. However
random HI motions effect the distance estimate.
Therefore, for a given input velocity we compute distances
corresponding to~$v\pm5\,$\kmsp. While errors can be quite large in
certain longitudinal directions, the distances in Table~\ref{t:results2}
have typical errors of $\sim0.5\ $kpc.

The error in the estimates of HI masses have contributions from the
distance estimates, the determination of the spatial extents of the
sources, the estimate of the number of channels associated with the
feature, and our subtraction (or non-subtraction) of the ambient
emission. The error in the distance ranges from $\sim 50\%$ at
distances $\lesssim 1\,$kpc to $\sim 20\%$ at distances of 5~kpc and
beyond.  The uncertainties in determining the spatial extent of the
source arise since it is difficult to distinguish an object from
surrounding clouds.  While one person's measurements may vary by $\sim
10-20\%$, the variation in this estimate between individuals (using
different software packages) is up to $25\%$.
The selection of a different range of channels results in a difference in
mean temperature of about 10\%.
The distance and area
measurements alone combine to give an uncertainty between 
$60\%$ and $40\%$.
Additionally, in many areas of the Galactic plane, it is virtually
impossible to find cloud-free area of significant size to compute the
contamination by the diffuse background. The background subtracted
HI masses are typically less than $50\%$ the uncorrected HI masses
in Table~\ref{t:results2}. Combining this with the uncertainties above
yields a total uncertainty of at least $80\%$ for nearby objects 
($\leq$ 1 kpc) and
$60\%$ for more distant objects.

\section{Results}
\label{s:results}

\notetoeditor{Please place Table~\ref{t:results} nearby.}

The objects found in our search, including the narrow
vertical structures, are listed in
Table~\ref{t:results} (see~\S~\ref{s:measure} for measurements).
Vertical objects in our list were designated using the format
``INVS Gllll+bbb+vvv". In anticipation that the more extensive
IGPS data  (\S~\ref{s:CGPS}) will contain vertical structures,
``INVS" stands for IGPS Narrow
Vertical Structures. ``Gllll+bbb+vvv" gives the location for the
extremity of the structure closest to the Galactic plane, including
its centroid velocity. Some irregular structures are worth
mentioning, and their locations are given as above without the
INVS designation. However, in the case of shells and cavities,
we give the Galactic coordinates of the center of the shell/cavity
along with its centroid velocity. See Table~\ref{t:results2} for
the derived properties of the structures and~\S~\ref{s:measure}
and~\S~\ref{s:errors} for analysis and uncertainties. Most of the
structures that are vertical are within 5~kpc from the Sun.

Although many structures were observed with almost
vertical orientations, no structure was found with a
morphology similar to \mushroom\ with or without the velocity
signatures of its stem and the cap. We also failed to detect
those objects in the KHR catalog which had a substantial portion
of their structure within the CGPS longitude and latitude
range (see~\S~\ref{s:KHR_null}). 

No clear narrow vertical structures of 500~pc height extending
between the Galactic disk and the halo were apparent in the
search in the appropriate velocity per Galactic
longitude\footnote{One other author also examined 4~fields
at the longitudes and velocities in which mushroom-shaped
clouds with heights of 500~pc should be visible and did not
recognize any.}, described in~\S~\ref{s:refined-search}.
Only 10~vertical
structures were found over $73\deg$~of longitude and 200~\kmss
of the CGPS. Five out of 10~of these have heights $\gtrsim 200$~pc.
We elaborate on the individual structures in~\S~\ref{s:indi}.

The vertical structures have no associated radio continuum
emission. Three of the vertical structures have 20-cm emission in the
field, however there is no morphological correlation between the radio
continuum and HI structures.  Eight of ten vertical structures have
detectable IR emission in the field of view associated with
their location, however only in five cases is there a significant
morphological correlation between the HI and IR emission.

We also studied the IR emission from \mushroom, and
constructed ratio maps (\IRR10060) using CGPS IR data to find new
mushroom-shaped structures.
One new candidate was discovered
using this method. INVS G130.6-3.1-36 is unremarkable in its
properties, and given its small size ($\lesssim100\,$pc), is
not likely important in the disk-halo interaction scenario.

The HI masses of the vertical structures fall in a wide range, from $\sim
100\,M_\odot$ to $\sim 2 \times 10^5\,M_\odot$. There are possibly
three distinct subgroups, $M\lesssim 10^3\,M_\odot$, $2 \times
10^4\,M_\odot \leq M\leq 6 \times 10^4\,M_\odot$, and $M >
10^5\,M_\odot$.

In the following we discuss some interesting structures
in more detail. None except for INVS~G111.4-0.2-45 appears to
be an important conduit in the disk-halo interaction.

\subsection{Searching for Footprints of Worms in the KHR Catalog}
\label{s:KHR_null}

Worm candidates in the KHR catalog (see \S~\ref{s:intro})
appear to extend away from the Galactic plane to the halo of
the Milky Way and a fraction of these were argued to be
the broken walls of HI surrounding burst superbubbles. KHR
found one third of their worm candidates in the outer Galaxy.
Apart from searching for new structures we also looked for
signatures of these worm candidates in the CGPS data. However
only seven KHR worms have $\gtrsim 50\%$ of their area within
the longitude-latitude range of the CGPS data (see
Table~\ref{t:KHR_null}). KHR determined the velocity of
each candidate by cross-correlating their final map of
candidates with channel maps integrated over ~4~\kmsp and
quoted a velocity range for each worm candidate. Within
its quoted range each candidate in the CGPS should display
a significant column density enhancement over a vertical
angular scale close to that cited in the catalog.
We also expect a worm to display a continuous velocity
signature along the components of its structure--- defined
by the column density enhancement--- since they have a
common origin. HI emission exists on all scales in the
Galaxy, and if not checked for velocity patterns, small
distinct clouds may be misidentified as parts of one
single large-scale structure. Therefore, we searched for
contiguous velocity signatures in the KHR worm candidates
using integrated HI brightness temperature maps and mean
velocity maps (\S~\ref{s:mom_maps}) and individual-channel
images.

Except for GW123.4-1.5, we could not confirm any of the
other 6~candidates as one continuous extended vertical
structure due to one or more of the following reasons: 
 
\begin{itemize}
	\item the integrated HI brightness temperature maps did not display a
significant column density enhancement that would be consistent with
a narrow vertical structure resembling the cataloged object at the
specified location.
	\item Individual channel images showed amorphous structures
on all spatial scales overlapping in a quasi-random fashion.
	\item Such amorphous structures displayed no continuity in
mean-velocity maps, which suggests that there is no systematic motion.
        \item The velocity fields of vertical structures apparent in
integrated HI brightness temperature maps were random. That is, they did not display
continuity when viewed channel by channel. For example, none of these
structures show a flow towards or away from the Galactic plane which we might
expect to see if the worm were inclined to our line of sight.
 
\end{itemize}

As an example,
we plot the integrated HI brightness temperature and mean-velocity maps
(\S~\ref{s:mom_maps}) of GW131.3-2.2 in Figure~\ref{f:KHR_null1}.
\notetoeditor{Please put Figures~\ref{f:KHR_null1}~a \&~b here.}
These maps were made over the entire velocity range specified by KHR,
\viz -59.6 \kmss --- -16.9 \kmsp. 
Figure~\ref{f:KHR_null1}~(a) shows the integrated HI brightness temperature map
in greyscale overplotted with contours of 100~$\mu$ emission;
the heavy line outlines the region of the worm specified in
the KHR catalog. 
It is quite apparent that the amorphous emission in the plane
dominates the image and there is little discernible enhancement in
HI and IR emission within the KHR outline of  GW131.3-2.2
(worm~\#$\,$56). Figure~\ref{f:KHR_null1}~(b) shows the mean-velocity map
in greyscale, the outline of the KHR worm candidate, and is overlaid
with  velocity contours
in order to bring out trends in the velocity map. The
velocity contours appear quasi-random which is 
suggestive of turbulent motion of the interstellar gas.  

The above approach is not sufficient since integrated maps present
mean trends over a large velocity range. That is, we would not expect
to recover a vertical structure like \mushroom\ with a
velocity half-width of $\sim 15\,$\kmss if it were embedded in the
velocity range given by the KHR catalog (\eg, for GW123.4-1.5, the KHR
velocity range reads $v=-58.3\,$\kmss to -10.8\kmsp).
Figures~\ref{f:KHR_null2} and~\ref{f:KHR_null3} describe another
approach. They display eight integrated HI brightness
temperature maps over the same area
of the sky as in Figure~\ref{f:KHR_null1}, each made by stepping
through the velocity channels in the specified velocity range of the
candidate. Each of these images was created by integrating six
neighboring channels, resulting in a velocity width of $\sim 5$~\kmsp.
A worm candidate should display a significant vertical column density
enhancement in more than one map, perhaps with only a part of the worm
existing in each image and  systematically appearing from one velocity to
another. Such a pattern could imply a vertical motion, but other
configurations of velocity structures were also included in
our search.

\notetoeditor{Please place Figures~\ref{f:KHR_null2} and~\ref{f:KHR_null3} facing each other nearby.}

Within the KHR velocity range of the worm candidate, there are
2~different density enhanced structures, each of which covers
2~integrated maps as required by our criteria. The first one
appears in Figure~\ref{f:KHR_null2}~(-31~\kmss $< v <$-22~\kmsp), but 
we note that the
velocity pattern seems more horizontal than vertical. The other enhancement
is shown in Figure~\ref{f:KHR_null3} (-46~\kmss $< v <$-37~\kmsp).
This feature is very diffuse and extends only to b$=-2.5\deg\ $
although the KHR catalog lists the  
worm candidate's total extent as b$=-4.0\deg$. Thus, neither
of these enhanced structures have the spatial or velocity
structure expected of a genuine HI structure. The portion of
the low-resolution
KHR worm that extends beyond $b = -3\deg\ $ is probably 
also composed of separate amorphous density enhancements
which are physically unrelated. We believe this since a
well-studied shell, GSH132.6--0.7--25.3 \citep[][]{nor00},
stands out clearly in the integrated HI brightness
temperature maps in Figure~\ref{f:KHR_null2}
(-31~\kmss $< v <$ -22~\kmsp), and the mean velocity map in
Figure~\ref{f:KHR_null1}. This demonstrates the 
usefulness of this approach
for detecting HI structures that have most of their area within the
latitude limits of the CGPS.

\subsection{Notes on Individual Candidates}
\label{s:indi}

We studied the vertical structures using the same techniques as for
KHR worm candidate search (see above) and describe the interesting
candidates in detail below.

\subsubsection{INVS~G79.3+0.9-39}

INVS~G79.3+0.9-39 was distinguished from the surrounding ambient
medium due to a cavity-like structure on its lower end and a velocity
continuity over~$\sim 12\,$\kmsp.  For a kinematic distance of
7.9~kpc \citep{bra93} , this structure has a height of 200~pc. Thus,
it is not a structure that may interact with the Galactic halo.  
We estimate a HI mass of $2.9 \times 10^3\,M_\odot$, much smaller than
the template worm \mushroom.  Figure~\ref{f:G79.4_low} shows that
well-correlated emission in 100$\,\mu$ and 21-cm radio continuum is
associated with an object displaced by $\sim20\arcmin$ away from
INVS~G79.3+0.9-39. We do not find convincing evidence of any
correlation between the vertical structure and this amorphous
emission. Interestingly, an HII region (SIMBAD ID: GAL 079.30+01.30)
at -10~\kmss relative to the central velocity of INVS~G79.3+0.9-39
appears in absorption along-side the HI vertical
structure. INVS~G79.3+0.9-39 may be associated with another vertical
structure INVS~G79.3+2.8-40, which appears at almost the same central
velocity (but not over the entire velocity range; see below).

\notetoeditor{Please spread Figures~\ref{f:G79.4_low}-~\ref{f:G84}
from here onwards.}

\subsubsection{INVS~G79.3+2.8-40}

INVS~G79.3+2.8-40 was identified in the HI cubes as
a faint extended vertical structure with a contiguous
velocity signature across its length.
At a kinematic distance of 7.9~kpc, this structure
starts at $\sim 200\,$pc from the plane and ends at
$\sim 500\,$pc away from the plane. Thus, this object
may be interacting with the halo of the Milky Way.
It has a total HI mass of $3.7 \times 10^3\,M_\odot$.
Figure~\ref{f:G79.3_high} displays the integrated
HI brightness temperature map for this structure
overplotted with contours of IRAS 100$\mu$ emission.
The candidate is the faintest of all our candidates,
and it appears in the same longitude as INVS~G79.3+0.9-39
with a large overlap in the velocity range. Both
structures are separated by an amorphous structure
in the middle~($b\sim 2\deg$) yet  a physical association
is suggested by the velocity overlap. In the event of
a possible association, the formation of such a structure
could be interesting from the point of view of the
disk-halo interaction scenarios.

\subsubsection{INVS~G111.4-0.2-45, G119.8+2.2-37,
G110.8-2.2-45: a candidate supershell?}
\label{s:supershell}

The individual-channel maps from the LRDS (see \S~\ref{s:CGPS})
are dominated by what seems to be 
one large-scale feature, which, at a kinematic
distance of 4.5~kpc, has a height of $\sim 500\,$pc. It has
the morphological appearance of a broken supershell. The velocity
signature, however, does not support this, and this object was
not previously cataloged 
as a supershell.
Figure~\ref{f:GSH115_NHI-ann} shows the integrated 
HI brightness temperature map of this
region made using the LRDS data combined with the 26-meter
single dish data acquired at higher latitudes for the CGPS-II
survey \citep{hig04}. This map
is integrated over the velocity range of v$=-55\,$\kmss to -30~\kmsp. 
This low-resolution structure contains four candidate vertical
structures: INVS~G111.4-0.2-45,
G119.8+2.2-37, G110.8-2.2-45, and INVS G115.2+1.9-41.
INVS~G111.4-0.2-45,
G119.8+2.2-37, and G110.8-2.2-45 are annotated in the figure,
showing that they extend to high latitudes.  INVS~G111.4-0.2-45,
 and G119.8+2.2-37 suggest the outline of a shell. INVS~G110.8-2.2-45
falls outside the region covered by the CGPS, and was identified
as a candidate based on the LRDS data alone. The fourth
structure, INVS G115.2+1.9-41, lays between the vertical
structures labeled in Figure~\ref{f:GSH115_NHI-ann} and is described
in more detail in a section below.
Figure~\ref{f:GSH115_mom0} displays the integrated 
HI brightness temperature
map of the shell-like structure, $ -3\deg <  b < 5\deg$, 
in the CGPS data,  integrated over the same velocity range as  above.

Figure~\ref{f:GSH115_COIR} shows the emission from
100$\,\mu$ (IRAS), CO (FCRAO) and HI (LRDS) over the same
portion of the sky, where the CO and HI data were both
integrated through the same velocity range specified
above. Though there is plenty of dust associated with
this region, the correlation between the dust and the
HI emission along the vertical extent of each subcomponent
labeled in Figure~\ref{f:GSH115_NHI-ann} is poor.
Therefore, these narrow vertical structures do not fit
the ``worm" definition (none of these structures were included
in the KHR catalog). The CO emission is concentrated in the
region of INVS~G111.4-0.2-45, and less prominently along 
INVS~G119.8+2.2-37. Given the overlap of molecular gas
in velocity and spatial coordinates with HI, we believe
a significant portion of molecular gas is associated
with the HI structure.

Six intensity maps, each integrated over 5~\kmsp, spanning
the velocity range above, are mosaicked in
Figure~\ref{f:GSH115_6maps}. INVS~G111.4-0.2-45 has prominent emission
over the entire velocity range, however, INVS~G119.8+2.2-37 is
prominent at velocities $v < -45\,$\kmsp.  There no signature of an
overall expansion in the velocity maps of these structures. In fact,
each vertical structure displays its own velocity behavior (or
spatial connections from channel to channel; \S~\ref{s:mom_maps})
which is distinct from others.
Given the different CO, IR, and high-resolution HI properties of
INVS~G111.4-0.2-45 and G119.8+2.2-37, they imply that these objects
are not walls of an expanding supershell.

Nevertheless, INVS~G111.4-0.2-45 is important from the point of view
of the disk-halo scenarios. It is $\sim 500\,$pc tall in CGPS-I and
the 26-m telescope HI data (30\arcmin \ resolution) from CGPS-II show
that it continues to maintain its structure at higher latitudes ($b >
6\deg$). Such a tall structure, and the presence of the associated
dust and molecular gas far away ($\gtrsim300\,$pc) from the plane
deserves a detailed study.

INVS~G110.8-2.2-45, for which we only have LRDS data, is displayed in
Figure~\ref{f:G112_mom1}. (Note that the HI mass estimate from this
lower resolution data  is an upper limit since it is likely to 
include neighboring clouds.) The figure shows dark contours of
integrated HI brightness temperature (over -37 to -51~\kmss)
overlapping the IRAS~100~$\mu$ emission map
represented in greyscale. There is fainter, yet significant
infrared emission extending below the LRDS range (b~$<-5\deg$,
following the light contours) and, from its morphology, we
believe that some of this emission could also be related
to the INVS candidate.

\subsubsection{INVS~G114.2-3.2-23}

Figure~\ref{f:G114_NHI} displays the column density map of
this INVS, with a blob-like structure \mbox{at b $\sim -3.2\deg$}
which narrows down along the longitudinal direction in the
lower part of the image (at \mbox{b $\sim -3.5\deg$}), appearing
like a narrow channel. At a distance of 2.2~kpc this object is
just 20~pc tall, but it has a very interesting velocity signature.
The mean velocity of the HI changes smoothly, yet significantly,
along the length of the narrow channel, and is portrayed in
figure~\ref{f:G114_mom1} (see \S~\ref{s:mom_maps}). The velocity
signature across the length displays a possible expansion
signature. Thus, the high resolution HI properties of this
object match those of the stem of
\mushroom\ (\S~\ref{s:template}). The extended structure on
top of the narrow structure ($b\sim 3\deg$) does not share
the same velocity signatures, nor does it show any distinct
correlated emission in the IR maps. Its association
with the narrow structure is not entirely clear (see below).

The archival LDS data show an extended HI structure at the location of
the candidate. There is discernible IR emission associated with this
structure (see Figure~\ref{f:G114_IRratio}), which has a narrow
vertical appearance with a ``bean"-like feature at its end. The
constructed IR ratio map using 100~and 25~$\mu$ IRAS images (obtained
from the SkyView database) do not distinguish the extended emission
from the ambient emission.  However, the HI and IR morphology of this
structure match reasonably with our template. This is a likely
candidate for another mushroom-shaped vertical structure, which needs
to be verified with high-resolution HI data at lower latitudes ($-4.5
< b < -3$). That may also clarify the entire morphology of this object
and its association with the extended structure on the top~(b $\sim
-3\deg$). Even with a likely extension to -4.2$\deg$,
INVS~G114.2-3.5-23 would be only~$\sim50\,$pc tall. Thus, it is
insignificant from the point of view of disk-halo phenomena.

\subsubsection{INVS~115.2+1.9-41}

Figure~\ref{f:G115.2_mom0IR} displays the integrated 
HI brightness temperature
map of this structure in greyscale overplotted with contours
of $100\mu$ emission. At a distance of 3.8~kpc, the object is
$\sim 200\,$pc tall. 
Even though there is a slight velocity
gradient across the length, it is not significant and our
velocity analysis may also be affected by the unavoidable inclusion
of ambient emission. 
The IR
emission displays some spatial correlation with the HI map
at the bottom of the structure (up to~b$\sim 3.4\deg$), which
reduces in the top half of the structure (up to~b$\sim 5\deg$),
where the HI structure appears fragmented . Faint IR emission
is found over a large angular area without any visible correlation
with HI morphology, it appears fragmented in the figure due
to chosen contour levels.
This object falls within the supershell-like
structure discussed in \S~\ref{s:supershell} and has a significant
overlap in velocity channels. However, its narrower velocity
spread makes INVS~G115.2-1.9-41 less conspicuous in
Figure~\ref{f:GSH115_NHI-ann}, which is made over a
velocity interval of~$> 30\,$\kmsp.  The structure is more
discernible in channel maps shown in Figure~\ref{f:GSH115_6maps}.
Given its height of 200~pc, it is not likely a significant
player in the disk-halo scenario.

\subsubsection{INVS~G76.8+0.8+20 and G85+1.6+33}

Figures~\ref{f:G78} and~\ref{f:G84} display two vertical structures of
large angular extent, both of which are $\sim 100\,$pc
tall. INVS~G76.8+0.8+20 and G85+1.6+33 appear at forbidden
velocities and do not display any signature of a systematic gradient
along their length.  Their positive velocities could be the result of the
typical 10-20 \kmss motion caused by the gravitational tug by the
Local Arm on these gas clouds. Also, given that these longitudes
intersect lines of sight through the inner part of the Galaxy, a
maximum red-shifted (positive) velocity would be measured at a tangent
point. If either of these objects is close to such a point plus in a
spiral arm, again they could display positive velocities.
Neither of them displays significant spatially
correlated emission in the IR, radio continuum, or CO maps.
While their morphology suggests a stellar wind origin,
it is really their small scales, that imply that they 
do not play a role in the
disk-halo scenario.

\section{Discussion}
\label{s:discuss}

In Table~\ref{t:results}, we list 10~candidate structures
matching our criteria of narrow vertical structures. Earlier
we discussed properties of the interesting ones
individually~(\S~\ref{s:indi}).

There is a range of physical sizes, morphologies, column
densities, and masses observed in these HI structures,
including the vertical structures. Most of the structures
are much too small ($\lesssim 100\,$pc) to be important in
the disk-halo connection.
However, there are five narrow vertical structures
which originate from the Galactic plane and reach to a height
of $\gtrsim 200\,$pc. 

Since INVS~G111.4-0.2-45 is $\gtrsim 500\,$pc tall, it 
may be involved in some mass and energy transport from the Galactic
plane.  However, this giant structure was not included in the KHR
catalog, even though it has the HI properties matching the worm
candidates. It was also not found when the KHR algorithm was utilized
to confirm candidates using the LDS data
\citep[][see~\S~\ref{s:intro}]{nor03}.  We believe that this structure
was missed due to a combination of factors: 1)~\citet{koo92} used
column density maps to produce the catalog; 2)~there is a
lot of structure in the HI channel maps around~$l=110\deg$ ; 
and 3)~the
IR emission does not correlate well with the morphology of
INVS~G111.4-0.2-45. The first two reasons are primarily responsible
for our non-confirmation of KHR candidates in the CGPS fields
(\S~\ref{s:KHR_null}). As detailed in \S~\ref{s:KHR_null}, all of the
six KHR worm candidates were found to be composed of smaller, unrelated HI
structures overlapping in the velocity maps. In order to avoid this
sort of confusion of features from unrelated structures, one has to
view channel maps with enough spatial and velocity resolution. The
mean velocity map can also be useful in such an exercise
(\S~\ref{s:mom_maps}). We caution that we could only search for six
KHR candidates in the CGPS data (apart from the confirmed worm
\mushroom), given the narrow latitudinal coverage of the
survey. However, our results show that we need to revise our ideas
about worm population, and data away from the Galactic plane will be
needed for a comprehensive understanding.

The 10 HI structures also differ widely in the amount of the
associated dust and molecular emission. Five structures with
length~$\gtrsim 2\deg$ have an IR morphology correlated
with their HI morphology. For two structures, at some locations
molecular emission in the same velocity channels was found to
to coincide with portions of the HI morphology, but not along
the whole extent. Thus, we did not find any systematic pattern
in the properties of the different objects.

\subsection{How Many Narrow Vertical Structures?}

Since we have conducted an unbiased search for structures with
high quality, uniform data, we can reasonably extrapolate to
estimate the total of number of observable narrow vertical
structures with similar characteristics in our Galaxy. Such an
estimate is useful to constrain various physical models of
the formation of these objects. 
However, our search had its own systematics and limitations, discussed
in~\S~\ref{s:systematics}, which the reader should keep it in mind
in~\S~\ref{s:numbers}.

\subsubsection{Systematics of Visual Inspection}
\label{s:systematics}
A visual search such as ours is the simplest approach for detecting
structures in HI cubes. However, it is subject to the analyst's
visualization scheme and the constraints imposed by the data. 

\notetoeditor{Please place Figures~\ref{f:local-NVS}(a \& b) here.}

Probably the most significant limitation is the latitude
range of the CGPS. Our intuition is that we do not recognize
vertical structures, since the ``footprint'' of one looks
quite different than the expected morphology of an~INVS.
Two examples, other than \mushroom\ itself, demonstrate this.
In Figures~\ref{f:local-NVS}(a \& b) we show the CGPS data range
for local arm velocities that display a broad amorphous
structure, indistinguishable from other ambient medium
features at b $\lesssim 0\deg$. At latitudes
beyond the survey limits (b $< -3\deg$), we see that the
structure is the broad
base of an INVS that has a channel much like the stem in
\mushroom. Another example is in Figure~\ref{f:GSH115_mom0},
which is a supermosaic of CGPS HI mosaics from fields MC-MF
on both sides of the Galactic plane (8~fields in total).
Examining individual fields alone did not reveal the full
extent of structures. 
Motivated by an apparent feature in the low-resolution LRDS
maps, we combined the CGPS fields
and the
resultant map shows that INVS~G111.4-0.2-45 is a vertical
structure with an angular height of~$\geq 5\deg$. Both
these examples indicate that structures extending to high
latitudes do not have definitive ``footprints" in the
plane, and that we have only recognized a fraction of INVS.

This does not explain why we do not detect disk-halo
conduits at distances larger than~8~kpc, however.
In~\S~\ref{s:refined-search}, we elaborate on our
refined search for objects with heights~$\gtrsim 500\,$pc.
We note, that even the limited latitude coverage and resolution
of the CGPS should allow the identification of 
a mushroom-shaped candidate,
between $\sim 8-23$ kpc.
Perhaps this rarity of INVS can be explained by a low column density contrast
between the feature and the ambient medium.

For example, the mean excess column density of the mushroom-shaped worm
GW123.4-1.5 above the background, over the velocity range
of the entire structure, is $N_{HI}=6\times10^{22}$cm$^{-2}$,
which is roughly equal to the average background over the same
velocity range. The entire structure is only discernible in data 
collected $\sim3\deg$ beyond the CGPS survey, yet even the stem appears
prominently in 
CGPS channel maps partly due to the fact that the worm is located
in the HI gap in LSR velocity between the local and the Perseus
arm. However,  similar candidates may not
be so fortuitously located and thus may not be visible due to
residing in a region 
of substantial ambient emission.
This makes a nonlinear scaling of the greyscale in a channel map
necessary to enhance weak extended features, and this scaling needs to
change from channel to channel depending upon the intensity in the
entire channel map. Although, our analysis consciously included a
variable scaling over the entire channel range, we still may not have
recognized vertical features which lacked a velocity gradient, or
other strong velocity signatures.

To access the dependence of our search on an individual analyst's
visual bias, we performed a limited test in which four of us
independently searched multiple CGPS HI mosaics and compared
our lists of detected objects. Although it would be valuable
to compare our assessment of vertical structures with computer
algorithms that detect structure in the CGPS HI cubes, at this stage
the algorithms are limited to finding HI small-scale or shell-like
phenomena associated with SNR, HII regions and Wolf-Rayet stars
\citep{dai03,kha04}.

The 3 co-authors analyzed animations of some mosaics of the CGPS
data cubes, particularly 2 mosaics which the first author assessed
as containing INVS. They searched for features which satisfied the
first 2 criteria in section 3.1. That is, they looked for
vertical objects with an appearance of spatial and velocity
continuity over 8 channels. Although an individual author did not
duplicate the detections of the first author, the combined
detections by the co-authors recovered all three INVS and the one
irregular structure in this field. Many other irregular structures
were found. However, subsequent assessment showed that their
motions were too random to consider the objects as genuinely
spatially connected. All authors detected the known shell
GSH132.6-07-25.3 demonstrating that the visual inspection
approach works well for an object that appears moderately cohesive
over ~10 km/s, even if it is juxtaposed on ambient emission
features. Significantly, none of the four analysts assessed
the emission apparently associated with the KHR worm candidate
GW131.3-2.2 (see \S~\ref{s:KHR_null}) as belonging to a narrow
vertical structure.

To summarize, this exercise demonstrates that the main analyst is
unlikely to mistake as INVS any structures produced by turbulent 
cloud motions. In fact, due to a strict consideration of a feature's 
velocity behavior, his number of selected features is lower than
the combined number of features selected by the other analysts.  
Also the number of resulting INVS is likely to be conservative due
to the low column density contrast or inability to recognize INVS
footprints. However each of his cataloged features has been 
confirmed by one other analyst. Therefore it is unlikely that the
main analyst missed a moderately cohesive, vertical structure that 
was recognized due to being narrow very near the midplane.

\subsubsection{The Galactic Population of INVS}
\label{s:numbers}
The CGPS survey
covers a longitude range of $74\deg-147\deg$, which is
roughly~20\% of the planar surface area of the Galaxy outside
the solar circle. Adopting an outer radius for  the Galaxy of
at least 18 kpc \citep{wol03} and a solar radius of 8.5~kpc,
the survey area amounts to 16\% of the total surface
area of the Milky Way. We found one  narrow vertical
structure with a vertical extent greater than 450~pc 
within the surveyed area of the plane. 
If we further assume such vertical structures
are distributed uniformly across the Galactic plane, we expect
the total number to be~6.  

Such a number is much smaller than the KHR catalog of the candidate
worms over the entire Galaxy. As we discussed above, many of the
structures in their survey could result from the confusion of HI
emission structures over many channels. The distribution of vertical
structures in the Galaxy is the only assumption made here. However, if
connected to star formation, these structures are expected to be
associated with the spiral arms and may be more frequent in the inner
Galaxy where the star forming rate is higher than in the outer Galaxy. Also,
we believe our detection rate to be low due to the limited latitude
range and other limitations associated with the visual inspection
described above.

\subsection{Possible origins of high-z structures}

As mentioned earlier, there could be three possible
subgroups of vertical structures based on their HI masses
($M\lesssim 10^3\,M_\odot$, $2 \times
10^4\,M_\odot \leq M\leq 6 \times 10^4\,M_\odot$, and $M >
10^5\,M_\odot$).
Given the apparent range in sizes and properties, we
believe that a single type of event could not give rise to
all these diverse vertical HI structures.

Worm candidates were postulated to be the remnant HI walls of
supershells associated with superbubbles. What can be said about the
vertical structures found in our search? 
Given a general lack of associated HII regions, none of our structures
appear to be related to supershell phenomena. 

Due to its small HI masses ($\sim100 M_\odot$)
and poor association with dust, we can speculate that G85+1.6+33,
could have been formed by shocks, say due to stellar winds or younger
SNRs.

Although INVS~G79.3+0.9-39 has a significant HI mass, it may also be an
example of a feature associated with a stellar wind phenomenon.
Displaced from its base is an HII region situated within an
HI cavity. There is significant radio continuum and
IR emission superimposed around this cavity-like structure
(designated in Table~\ref{t:results} as G79.4+1.1-41). 
INVS G79.3+2.8-40 could be an extension of INVS G79.3+0.9-39,
if so we can speculate that they resulted from past stellar
wind activity associated with the HII region.

INVS~G111.4-0.2-45 appears to be large enough to be important in the
disk-halo scenario. For comparison, the CGPS HI chimney, or burst
superbubble, candidate \citep[discovered by][]{nor00} located above
W4 star forming region, has IR and radio continuum emission that is
associated with the HI structure appropriate for a
chimney. INVS~G111.4-0.2-45, on the other hand, does not have a large
star forming region near its base. Additionally, it is hard to explain
the molecular gas apparently associated with the HI at~$4\deg$ ($\gtrsim
300\,$pc) away from the plane. Thus, an association of this structure
with  chimney-like activity seems unlikely.

Alternatively, we could imagine,
that INVS~G111.4-0.2-45 resulted from a collision
of a high velocity cloud with the Galactic plane
\citep{ten87}. Formation of
large structures with large HI masses ($10^4-10^6\,M_\odot$)
is possible in this scenario. The low shock velocity
in such collisions generates much less ionized gas
than the chimney activity, and self gravity in
the cooled, shocked gas can result in clouds of molecular
gas far away from the plane. A cloud collision can take
place anywhere in the plane of the disk, which makes
an association with a star forming region unnecessary.

However the scenarios most relevant to our investigation are the
ones that
form extended disk-halo features by means of buoyant forces. In these
scenarios, a single SN need not be the only source of input energy for
the creation of buoyant structures; it can be provided by the ionized
gas reservoirs \citep[][and references therein]{dea01}, a
collision between clouds, or even the modest impact of a cloud falling
onto the plane.  Although in the CGPS-I it lacks the mushroom-shape,
INVS~G111.4-0.2-45 has a comparable HI mass and extent to \mushroom\ and
so we should not neglect the buoyant cloud picture.  The
preliminary examination in CGPS-II which suggests it may extend to
twice this height does not eliminate the buoyant structure scenario,
since \cite{irw03} show that structures with $10^7 M_\odot$ can
theoretically evolve buoyantly for many kilo-parsecs; their detailed
model reproduces an observed HI disk-halo feature in an external
galaxy. Additionally a buoyant feature will
have a mushroom-shape only for part of its evolution and will 
appear simply as a vertically extension, particularly at late
evolutionary stages. We have examined existing simulations by
de~Avillez and his collaborators, where features similar to
but a little shorter than INVS~G111.4-0.2-45 are created by
stellar activity in the disk, surviving over a few million
years. Specifically, the one in \citet{dea00}'s figure~13(b)
at position (200, -300~pc) is $\lesssim 400$~pc tall.
This simulation is the same in which \citet{dea01} find
buoyant mushroom-shaped clouds. Independently, simulations
by \citet{kor99} also show similar structures, see their
Figure~5 (at 250, 500~pc), surviving over~2-3~Myr.

\subsection{Mushroom-shaped INVS and old SNRs}
\label{s:mushroom-oldSNR}

Finally, what can we say about mushroom-shaped worms?
Since we did not detect any more vertical structures
with that morphology, we are not in a position to
establish the evolution of old SNRs into a mushroom
shaped cloud \citep{eng00}. However, assuming this
scenario occurs, we can consider the rarity of
this observation. 

In the toy model produced for \cite{eng00} an evolving
SNR develops a mushroom-shaped cap that lasts for a
couple of Myr. Assuming a supernova rate of 1~per
50~years, we will have $\sim 4 \times 10^4$ SNRs in
the entire Galaxy over this period and $\sim 6 \times 10^3$
may have formed over 16\%~of the disk surface area.
Thus we observe one buoyant cloud out of the several
thousand that could potentially form mushroom-shapes.

While the issues related to visually detecting HI clouds
(\S~\ref{s:systematics}) play a role in producing the
limited number, so could the environment of the SN blast.  
Firstly, the SN must occur in a low density region in
order to evolve for a number of Myrs into a mushroom
shape. Secondly, a single SN needs to be at least
slightly displaced from the midplane in order to
experience the pressure gradient that will cause
the bubble to rise buoyantly.  Its rise may be less
inhibited if the SN event occurs higher out of the
midplane or between spiral arms. While these
displacements may also render the cloud less confused
with ambient emission, the cloud may not accumulate
a sufficient amount of HI gas and/or dust during its
evolution to be detectable. Alternatively, if a rising
buoyant cloud encounters a dense medium, it may be
dispersed or become deformed beyond recognition.

In the supernova scenario, one may expect to find some similarities,
and even a systematic pattern in the properties of \mushroom-like HI
clouds and old SNRs. The difficulty lies in the paucity of old SNRs;
there are only 7 candidates with ages $\geq 10^5\,$yr in the
literature. Also, while it could be enlightening to compare the IR
properties of \mushroom-like structures with old SNR, only two of the
literature's old SNR candidates have morphologies that are distinct in
the IR (G166.2+2.5 and GSH132.6--0.7--25.3).  We found, following the
procedure of \citet{sak92} for forming IR ratio maps, that the
morphology of G166.2+2.5 is only apparent in \IRR10060 and only
marginally distinct from the ambient emission for GSH132.6--0.7--25.3
in \IRR10060. Thus the characteristic range of IR ratio values for a
normal population of old SNRs is unknown and we cannot currently use
this to property to determine if \mushroom\ is related to SNRs.
However this could be a strong test if larger IR and HI databases for
evolved SNRs existed; it would be useful if such an IR database avoids
the mid-plane, where it is difficult to distinguish shells from
ambient emission. In the absence of large databases, theoretical
models of the predicted IR and HI emission from old ($> 1$~Myr) SNRs
could be generated in order to support this test.

UV cooling is expected to be inefficient and hence
UV observations may be better for characterizing and
comparing old SNRs. However the intervening HI obscures
any potential UV radiation from \mushroom.

\subsection{Future Work}

Developing computer algorithms to detect vertical features
could circumvent the biases inherent in visual inspection. 
More importantly, HI studies at higher
latitudes are needed  to securely recognize features important in the
disk-halo connection either visually or algorithmically. 
Currently, observations at
Galactic latitudes up to~$b = 18\deg$ ($110\deg < l 
< 118\deg$) in the second phase of the CGPS (CGPS-II),
are over. One of us (AA) is pursuing more
observations with the DRAO synthesis telescope at
latitudes of~$\gtrsim 6\deg$, contiguous to the
above mentioned data. Similar high-latitude synthesis
observations using the DRAO telescope have been
proposed to study the Perseus complex, which will
doubtlessly contribute to the disk-halo studies as well.

Future surveys, using instruments such as  Arecibo L-Band Feed Array
(ALFA), plan to map the sky in radio recombination lines in order to
detect the diffuse ionized gas generated by the disk-halo phenomena\footnote
{\url{http://alfa.naic.edu/galactic/recomb/alfa\_recomb.html}}. In the
case of \mushroom, rather than studying it in IR emission,
attempts could be made to detect recombination lines from any ionized
gas inside the cap. 
If gas with temperatures greater than $10^5\,$K were detected at
particular locations within the cap, this could support a SN origin
for this mushroom-shaped cloud.

\acknowledgments

This research is supported by the International Galactic Plane Survey
(IGPS), the Natural Sciences and Engineering Research Council (NSERC)
of Canada, and the University of Manitoba.  AA was supported by an
IGPS Postdoctoral Fellowship.  The CGPS is a Canadian project with
international partners, supported by a grant from NSERC. We are
grateful to J. Dickey, T. Landecker, S. Basu and T. Kudoh for useful
discussions. We also thank the anonymous referee whose suggestions
have improved our manuscript.

\clearpage


\begin{figure}
\caption{Velocity signature of \mushroom: (a,b,c) The integrated HI
brightness temperature over the LSR velocity ranges of -32 to
-41~\kmsp, -38 to -47~\kmsp, and -44 to -51~\kmsp, respectively; the gray
scale stretch is arbitrary.  The figures highlight the appearance and
velocity signature, in a velocity cube, of this worm which we used
to develop our
search template.  The stem appears from one velocity extremity to
another gradually, and the cap exists at velocities distinct from the
base of the stem.  (d) Mean-velocity map of the worm over -32 to
-51~\kmsp. An arbitrary stretch was used to highlight the velocity
signatures of the stem and the cap with respect to the ambient gas.
Blue was assigned to the more red-shifted velocities.  The stem has a
marked velocity gradient of $12\pm 3$\ \kmss along its length and a
signature of a possible expansion across its width. The cap does not
show signatures of any significant expansion, although the lobes
appear to be falling towards the Galactic plane. Note that we do not
include the structures at latitudes $b < -5.0\deg$ as part of the cap.
\label{f:mush-vel-sig}}
\end{figure}

\begin{figure}
\caption{Infrared-ratio map (\ir10025) of \mushroom. This ratio
map was made from background subtracted infrared images from IRAS
survey; the greyscale stretches over values of 12-37, where
37~is black. The contours of an HI integrated brightness temperature
map (300~to 900~K, in steps of 300K), over the velocity range
of -49.5 to -25.8 \kmss and smoothed over 5 pixels, are
overplotted to indicate the outline of the worm structure. The cap is
well demarcated; however the stem remains undistinguished in either
the ratio map or the individual IR maps (not shown).  Note that we
do not include the structures at latitudes $b < -5.0\deg$ as part
of the cap.
\label{f:mush_IR100-25}}
\end{figure}

\clearpage 

\begin{figure}
\caption{{\bf(a)} Integrated HI brightness temperature map of the worm
candidate GW131.3-2.2 overplotted with contours of IR 100-$\mu$ emission
(10 to 70~MJy~Sr$^{-1}$, where MJy is megaJansky). The CGPS velocity
range of  the H~I greyscale is 17.1 \kmss to -59.2 \kmsp.
An approximate hook-shaped boundary for GW131.3-2.2 (KHR's ``worm 56")
is drawn with a heavy line in black, and the worm is emphasized with dashed
lines. Another nearby KHR-worm candidate (\# 54) is also
delineated with KHR's boundaries for this object.
The region of $-1\deg < b < 1\deg$ was omitted in KHR's analysis: a heavy
line in white marks $b=-1\deg$. {\bf (b)}~Mean-velocity map of the worm
overplotted with velocity contours between -34 to -44~\kmsp, in steps of
4~\kmss each. An approximate boundary of the worm candidates is marked as
in figure on the left.  
\label{f:KHR_null1}}
\end{figure}

\clearpage

\begin{figure}
\epsscale{0.9}
\caption{Integrated HI brightness temperature maps of GW131.3-2.2
over the velocity range $v = -17$~\kmss to -36~\kmsp.
An outline of the worm candidates is also drawn as
in figure~\ref{f:KHR_null1}. This and the next figure demonstrate
that, at the resolution of the CGPS,  one cohesive object doesn't span 
the spatial and velocity domain described in KHR.  
\label{f:KHR_null2}}
\end{figure}

\begin{figure}
\epsscale{0.9}
\caption{Integrated HI brightness temperature maps of GW131.3-2.2
over the velocity range $v = -37$~\kmss to -56\ \kmsp.
An outline of the worm candidates is also drawn as
in figure~\ref{f:KHR_null1}.
\label{f:KHR_null3}}
\end{figure}

\clearpage

\begin{figure}
\caption{Integrated HI brightness temperature map of INVS~G79.3+0.9-39
over the velocity range from~-30 to~-46~\kmsp. The greyscale
in the figure stretches from about~200 (white) to~1600~K (black).
The vertical structure with a width of $l=79.1\deg$ to
$l=79.2\deg$ is at the center of the map, extending vertically
from a latitude of $b=0.8\deg$ to $b = 2.4\deg$, with a
longitudinally extended diffuse structure at $b\sim2\deg$.
An HII region at the location of $l\sim 79.3\deg$ and
$b\sim 1.3\deg$ appears in absorption. (Note, 1$\deg\ $ is
equivalent to 135 pc.) We have
overplotted black contours of IRAS 100~$\mu$ emission
from~700 to 3000~MJy~Sr$^{-1}$ in steps of 1000~MJy~Sr$^{-1}$,
and white contours of CGPS radio continuum emission at
21~cm in the range 20-50~K
in steps of 15~K. There is significant IR and radio
continuum emission, which are well-correlated with
each other but not
with the HI map of the vertical structure.
\label{f:G79.4_low}}
\end{figure}

\clearpage

\begin{figure}
\caption{Integrated HI brightness temperature map of INVS~G79.3+2.8-40
over the velocity range from~-36 to~-49~\kmsp.
To emphasize the outline of the structure, we have
overplotted black contours of the same map in the
range 350-550~K, in steps of 150~K. We have overplotted
white contours of IRAS 100~$\mu$ emission from
200-1000~MJy~Sr$^{-1}$ in steps of 200~MJy~Sr$^{-1}$. There is significant
IR emission, though it does not show convincing spatial correlation
with the HI map.
\label{f:G79.3_high}}
\end{figure}

\clearpage

\begin{figure}
\caption{Integrated brightness temperature map from LRDS covering the
longitude range $105\deg < l < 125\deg$. This map was made by using
LRDS data (\S~\ref{s:CGPS}) from -30~\kmss to -55\ \kmsp; for a velocity
of -47 \kmss at these Galactic longitudes, $1\deg \sim 77\,$pc. 
This also includes some recently acquired observations for CGPS-II
high-latitude fields by the DRAO 26-meter dish \citep{hig04}.
The greyscale goes from 0~K to 1700~K, and
the contours are overlaid to highlight faint structures far away from
the Galactic plane. Three candidate vertical structures are annotated
along with the location of Cas-A SNR, which appears in
absorption. The high resolution CGPS mosaics do not cover
the entire area of INVS 110.8-2.2-45, and it was studied
using LRDS data alone.
\label{f:GSH115_NHI-ann}}
\end{figure}

\clearpage

\begin{figure}
\caption{Integrated HI brightness temperature map of the
region between the Galactic longitudes of 107$\deg$ and 123$\deg$. This
map was constructed over the velocity of $v = -30$\ \kmss to $-55$\ \kmss
made using the CGPS data, where the greyscale ranges from 0~K to 3300~K.
Two candidate vertical structures INVS~G111.4-0.2-45 and
G119.8+2.2-37 can be seen. There is a (white) gap in this map in the
region that includes Cas-A SNR ($l\sim 112\deg$, $b\sim -2.2\deg$)
since it is difficult to remove artifacts from images near bright radio
sources.
\label{f:GSH115_mom0}}
\end{figure}

\clearpage

\epsscale{1.0}
\begin{figure}
\caption{IRAS 100$\mu$ map of the region between the
Galactic longitude of 107$\deg$ and 123$\deg$. The greyscale
spans from 19~MJy~Sr$^{-1}$ to 9000~MJy~Sr$^{-1}$. An  outline
of the integrated HI brightness temperature map derived from the LRDS
(\S~\ref{s:CGPS}) data over the velocity range
\mbox{v$=-30$\ \kmss to $-55$\ \kmss} is plotted using thick
white contours. The black contours represent the integrated
HI brightness temperature from the CO spectral-line data
over the same velocity range (5~to 50~K in steps of
5~K). Note that a significant amount of molecular gas
and dust appear to be associated with the two vertical
structure candidates, and are present far away
($\gtrsim300\,$pc) from the plane of the Galaxy.
\label{f:GSH115_COIR}}
\end{figure}

\clearpage

\epsscale{0.85}
\begin{figure}
\begin{center}
\caption{High-resolution channel maps of the region
between the Galactic longitudes~107$\deg$ and~123$\deg$.
Six figures show the integrated HI brightness temperature maps
of the candidate vertical structures between -24\ \kmss to 53\ \kmss
over 5~\kmss each. The greyscale values range from 0~to about
400~K. A contour outline of the integrated HI brightness temperature map of
this region made using the low-resolution LRDS data is overlaid.
\label{f:GSH115_6maps}}
\end{center}
\end{figure}

\clearpage

\begin{figure}
\begin{center}
\begin{tabular}{ccc}
\epsscale{0.3}
\end{tabular}
\end{center}
\caption{
	The candidate INVS 110.8-2.2-45 using LRDS and
	IRAS data. The figure shows the IRAS $100 \mu$
	emission, where the greyscale is stretched between
	20-160~MJy Sr$^{-1}$.
	The dark contours outlining the INVS morphology
	represent the HI brightness temperature map made by
	integrating the low resolution LRDS data over
	-51 $< v < $ -37 \kmsp (the contour levels
	are 700-1200 K in steps of 80~K.) Light grey contours
	(30-50~MJy Sr$^{-1}$ in steps of 10~MJy Sr$^{-1}$)
	are used to outline faint IR emission that appears
	to extend below the LRDS outline of the INVS.
        \mbox{($1 \deg\ $ = 77 pc.)}
\label{f:G112_mom1}}
\end{figure}

\clearpage

\epsscale{1.0}
\begin{figure}
\caption{Integrated HI brightness temperature map of
INVS~G114.2-3.2-23 over the velocity
range \mbox{v = -16~\kmss to -26 \kmss} using the CGPS
data. The greyscale here stretches over~109-571~K. The contours
represent the same map smoothed over~$\sim 5\arcmin$,
with levels from~280 to 380~K in steps of~30~K. This
structure consists of a blob-like structure
\mbox{at $b\sim -3.2\deg$}, which narrows down along the
longitudinal direction in the lower part of the
image at $b\sim -3.5\deg$. ($1\deg\ $ = 38 pc.)
\label{f:G114_NHI}}
\end{figure}

\clearpage

\begin{figure}
\caption{Mean velocity map of candidate INVS~G114.2-3.2-23 
between -16 to -25 \kmsp.  The
intensity weighted mean velocity map shows the velocity structure of
the candidate. This candidate displays a smooth gradient along the
narrow part ($b\lesssim -3.3\deg$), as well as a discernible gradient
across its width (latitudinal).  This velocity behavior is similar to
that of the template stem of
\mushroom.  (Blue is assigned to -16 \kmss and red to -25 \kmsp.)
\label{f:G114_mom1}}
\end{figure}

\clearpage

\clearpage

\begin{figure}
\epsscale{0.65}
\caption{IR emission possibly associated with
INVS~G114.2-3.2-23. The IRAS 100$\mu$ image is overlaid with
white contours of the integrated 
HI brightness temperature map derived from
the smoothened CGPS data (300-500 K in steps of~75~K).
The 100$\mu$ image greyscale image stretches
	between 12 (white) to 800 (black) MJy Sr$^{-1}$. 
\label{f:G114_IRratio}}
\end{figure}

\clearpage

\begin{figure}
\epsscale{0.55}
\caption{Integrated HI brightness temperature image of INVS~G115.2+1.9-41.
This CGPS map was constructed over velocities from~-33
to~-43~\kmsp, and is overplotted with black contours of IRAS
100$\mu$ images of the range of 75-100~mJy~Sr$^{-1}$ in steps of
15~mJy~Sr$^{-1}$.  The extent
from b=1.9$\deg\ $ to b=5.2$\deg\ $  corresponds to a linear scale of
200 pc and the widest portion at b=2.54$\deg\ $ is 24 pc. 
\label{f:G115.2_mom0IR}}
\end{figure}

\clearpage

\begin{figure}
\caption{Integrated HI brightness temperature map of INVS G76.8+0.8+20.
This map was made 
over a velocity range of 15 to 26\ \kmsp, where the greyscale 
extends from 29K~to 689~K. The overplotted contours represent the
IR 100$\mu$~emission ranging between 300~and 1200~mJy~Sr$^{-1}$,
in steps of 300~mJy~Sr$^{-1}$. The extent
from b=0.8$\deg\ $ to b=3.6$\deg\ $  corresponds to a linear scale of
95 pc and the widest portion at b=3$\deg\ $ is 28 pc.
\label{f:G78}}
\end{figure}

\clearpage

\begin{figure}
\caption{Integrated HI brightness temperature map of INVS G85+1.6+33. This map was made
over a velocity range of 23--40\ \kmsp, where the greyscale
extends from 0~to 300~K. The overplotted contours represent the
IR 100$\mu$~emission ranging between 150~and 400~mJy~Sr$^{-1}$
in steps of 100~mJy~Sr$^{-1}$. The extent
from b=1.6$\deg\ $ to b=4.6$\deg\ $  corresponds to a linear scale of
40 pc and the widest portion at b=1.95$\deg\ $ is 6 pc.
\label{f:G84}}
\end{figure}

\clearpage

\begin{figure}
\caption{We use these two figures of the same stem-like object to
illustrate, that in the survey latitude range of~$-3\deg < b < 0\deg$
the footprint of a narrow vertical structure can be indistinguishable
from the ambient medium. The vertical structure becomes apparent at
lower latitudes and exists in 6 continuous channels.  Each of the
figures shows the sum of 2~channels with the Figure~(a) starting
at~$\sim$ -8 \ \kmsp. Recall each channel is 0.82 \kmsp wide.
\label{f:local-NVS}}
\end{figure}

\clearpage






\newpage

\begin{deluxetable}{ccccccc}
\tabletypesize{\scriptsize}
\tablecaption{List of narrow, vertical structures and 
serendipitously found objects. Although we did not search rigorously
for shells, cavities, and amorphous clouds, this list includes 
interesting objects which caught our attention and which may 
not have previously been studied.\label{t:results}}
\tablehead{
\colhead{Name\tablenotemark{a}} & \colhead{CGPS\tablenotemark{b}}
& \colhead{Height\tablenotemark{c}} & \colhead{Width\tablenotemark{c}}
& \colhead{Velocity\tablenotemark{d}}  & \colhead{column}\\
\nodata&\colhead{Field}&\colhead{(degrees)}&\colhead{(degrees)}&
\colhead{ FWHM (\kmsp)}&\colhead{density contrast\tablenotemark{e}}
&\colhead{Remarks\tablenotemark{f}}
}
\startdata
INVS G76.8+0.8+20&MO2&2.8&$\sim 0.75$&$\sim 6$& $>$ 1:2\\
INVS G79.3+0.9-39&MN2&1.5&$\sim0.16$&$\sim 12$& $<$ 1:1.5\\
INVS G79.3+2.8-40&MN2&2.25&0.3&$\sim10$& $<$ 1:3\\
INVS G85+1.6+33&MM2&3&0.16&$\lesssim5$&1:3\\
INVS G110.8-2.2-45&ME2&5.5&$\sim1.25$&15& $>$ 1:2\\
INVS G111.4-0.2-45&ME1&3&$\gtrsim 1$&$\gtrsim 12$&1:1.5\\
INVS G114.2-3.2-23&ME1&$>$ 0.5& 0.2 & 7 & 1:3\\
INVS G115.2+1.9-41&MD2&3&$\lesssim 1$&$\sim10$&1:4\\
INVS G119.8+2.2-37&MC2&2.5&0.5&$\sim 10$&1:2\\
INVS G130.6-3.1-36\tablenotemark{g}&MA1&1.3&0.25&6&$<$ 1:2&\\
&&&&&&\\
G106-1.5-128&MG1&2&2&$\lesssim 4$&$\lesssim 1:3$&shell/cavity\\
G116.1-2.3-110&MD1&2&2&5&1:5&candidate shell\\
G79.4+1.1-41&MN1&0.75&0.75&15& $<$ 1:1.5 & shell/cavity\\
G126.6+0.8-36&MA2&1&1&8&$<$ 1:1.5&irregular\\
G129.3+0.7-34&MA2&2&2&8&1:1.5&irregular\\
G129.7+1.3-37&MB2&1.5&1.25&10&$<$ 1:2&irregular\\
G146+2.5-40&MV2&1&0.3&8&1:1.5&irregular\\
\enddata
\tablenotetext{a}{See \S~\ref{s:results} for the convention
used to designate the objects. Note that the LSR systemic velocity
is the suffix in each name.}
\tablenotetext{b}{See \citet[][]{tay03} for the definition of the
CGPS field locations.}
\tablenotetext{c}{If a structure tilts east or west 
we quote approximate latitudinal difference
between the top and bottom of the structure as its ``height". The
widths of structures generally vary along their length; quoted are
the typical values.}
\tablenotetext{d}{The HI emission from the structure of interest
often has background emission in its wings. Estimating FWHM can be
problematic in such cases, and a visual estimate given here has a
typical error of~2~\kmsp. }
\tablenotetext{e}{Estimated from integrated HI brightness temperature maps.}
\tablenotetext{f}{Objects with a morphology unlike a shell or
a vertical structure were termed as `irregular'.}
\tablenotetext{g}{Located initially using the CGPS infrared maps.}
\end{deluxetable}

\newpage

\begin{deluxetable}{ccccccccc}
\tabletypesize{\scriptsize}
\tablewidth{0pt}
\tablecaption{Derived properties of narrow vertical structures listed in Table~\ref{t:results}\label{t:results2}}
\tablehead{
\colhead{Name}&\colhead{distance}&\colhead{height}&\colhead{width}&
\colhead{column density}&\colhead{HI mass}&\colhead{background}&\colhead{Corrected}\\
\nodata&\colhead{\& error (kpc)}&\colhead{(pc)}&\colhead{(pc)}&
\colhead{$10^{21}\,$cm$^{-2}$}&
\colhead{($M\odot$)} &\colhead{HI mass ($M\odot$)} & \colhead{Mass ($M\odot$)}
}
\startdata
INVS G76.8+0.8+20&$1.94$\tablenotemark{a}&95&25&0.516&$2.17\times10^3$&$1.4\times10^3$&$7.7\times10^2$\\
INVS G79.3+0.9-39&$7.7_{-0.5}^{+0.48}$&202&22&1.89&$1.29\times10^5$&$0.957\times10^5$&$3.3\times10^4$\\
INVS G79.3+2.8-40&$7.8_{-0.5}^{+0.42}$&306&41&0.709&$1.04\times10^5$&$4.72\times10^3$&$5.7\times10^4$\\
INVS G85+1.6+33&0.75\tablenotemark{a}&40&2&0.206&$1.91\times10^2$&$6.44\times10^1$&$1.3\times10^2$\\
INVS G110.8-2.2-45\tablenotemark{b}&$4.4_{-0.5}^{+0.5}$&230&77&1.66&$5.96\times10^5$&$3.59\times10^5$&$2.4\times10^5$\\
INVS G111.4-0.2-45&$4.4_{-0.5}^{+0.5}$&421&96&2.98&$3.92\times10^5$&$2.31\times10^5$&$1.6\times10^5$\\
INVS G114.2-3.2-23\tablenotemark{c}&$2.19_{-0.46}^{+0.43}$&19&8&0.7&$1.20\times10^3$&$6.55\times10^2$&
$5.5\times10^2$\\
INVS G115.2+1.9-41&$3.47_{-0.47}^{+0.42}$&182&60&1.04&$6.82\times10^4$&$4.23\times10^4$&$2.6\times10^4$\\
INVS G119.8+2.2-37&$3.35_{-0.48}^{+0.42}$&146&30&1.26&$5.47\times10^4$&$3.34\times10^4$&$2.1\times10^4$\\
INVS G130.6-3.1-36&$3.1_{-0.5}^{+0.5}$&70&14&0.639&$6.98\times10^3$&$3.82\times10^3$&$3.1\times10^3$\\
\enddata
\tablenotetext{a}{For structures at forbidden velocities, the quoted
values are for distances to the tangent point locations.}
\tablenotetext{b}{Mass and error estimates from LRDS data only.}
\tablenotetext{c}{Only partially visible in the CGPS data.}
\tablecomments{See~\S~\ref{s:measure} for the procedure used
to derive the tabulated properties.}
\end{deluxetable}

\newpage

\begin{deluxetable}{cccccc}
\tabletypesize{\scriptsize}
\tablewidth{0pt}
\tablecaption{Parameters of worm candidates in the KHR catalog
that fall in the CGPS Sky.\label{t:KHR_null}}
\tablehead{ \colhead{Name} & \colhead{CGPS}&
\colhead{Central} &
\colhead{Velocity} \\
\nodata&\colhead{Field}& \colhead{Velocity (\kmsp)} &\colhead{width (\kmsp)}
}
\startdata
G89.7-2.4&ML1&-28.5&12.4\\
G93.7+3.0&MK2&-9.05&12.35\\
G99.4-1.7&MH1&-9.7&2.4\\
G107.4-3.3&MF1&-54.95&4.25\\
G123.4-1.5&MB1&-34.55&23.75\\
G129.8+1.6&MA2&-40.35&14.15\\
G131.3-2.2&MY1&-38.25&21.35\\
\enddata
\tablecomments{Except for GW123.4-1.5 (the Mushroom), none of
these candidates could be confirmed using the high-resolution data
from the CGPS, see~\S~\ref{s:KHR_null}.}
\end{deluxetable}

\clearpage


\begin{deluxetable}{ccccc}
\tabletypesize{\scriptsize}
\tablewidth{0pt}
\tablecaption{The parameters for a refined search
(\S~\ref{s:refined-search}). Velocities $v_{500}$
are given by the Galactic rotation model \citep{bra93} for the
locations at which a structure at a height from the Galactic
plane of 500~pc is visible in the CGPS data. $v_{stem}$
($v_{cap}$) is the velocity at which the stem (cap)
becomes inadequately resolved for a secure identification.
\label{t:refined-search}}
\tablehead{\colhead{CGPS field} & \colhead{Central longitude} & \colhead{$v_{500}$} & \colhead{$v_{stem}$} & \colhead{$v_{cap}$} \\
\nodata&\colhead{degrees}&\colhead{\kmss}&\colhead{\kmss}&\colhead{\kmss}
}
\startdata
MV & 144.7&-64&-91&-113\\
MW & 140.7&-70&-100&-124\\
MX & 136.7&-74&-108&-134\\
MY & 132.7&-78&-115&-143\\
MA & 128.7&-82&-121&-152\\
MB & 124.7&-84&-127&-160\\
MC & 120.7&-87&-132&-167\\
MD & 116.7&-88&-136&-174\\
ME & 112.7&-88&-140&-179\\
MF & 108.7&-88&-142&-184\\
MG & 104.7&-86&-144&-187\\
MH & 100.7&-84&-145&-190\\
MIJ& 096.7&-80&-145&-192\\
MK &  92.7&-77&-144&-193\\
ML &  88.7&-72&-143&-193\\
MM &  84.7&-66&-140&-192\\
MN &  80.7&-60&-137&-190\\
MO &  76.7&-52&-133&-187\\
\enddata
\end{deluxetable}







\begin{thebibliography}{}
\bibitem[de Avillez \& Mac Low (2001)]{dea01} de Avillez, M., \& Mac
Low, M.-M. 2001, \apjl, 551, L57
\bibitem[de Avillez (2000)]{dea00} de Avillez, M.~A.\ 2000, \mnras, 315, 479
\bibitem[Belfort \& Crovisier(1984)]{bel84} Belfort, P.~\&
Crovisier,~J.\~1984, \aap, 136, 368 
\bibitem[Brand \& Blitz (1993)]{bra93} Brand, J., \& Blitz, L. 1993,
\aa, 275, 67
\bibitem[Butler Burton (1988)]{but88} Burton, W. B. 1988a, in Galactic
and Extragalactic Radio Astronomy, ed. G. L. Verschuur \& K. I. Kellermann
(New York: Springer), 295
\bibitem[Burton \& Hartmann(1994)]{bur94} 
	Burton, W.~B.~\& Hartmann, D.\ 1994, \apss, 217, 189
\bibitem[Cao \etal~(1997)]{cao97} Cao, Y., Terebey, S., Prince, T. A.,
\& Beichman, C. A. 1997, ApJS, 111, 387
\bibitem[Daigle \etal (2003)]{dai03} 
Daigle, A.,  Joncas, G., Parizeau, M., \& Miville-Deschênes, M.-A.
2003, \pasp, 115, 662
\bibitem[Dennison \etal (1997)]{den97} Dennison, B., Topasna, G. A., \&
Simoneti, J. H. 1997, \apjl, 474, L31
\bibitem[Dickey \etal (2001)]{dic01} Dickey, John M., McClure-Griffiths,
N. M., Stanimirovic, Snezana, Gaensler, B. M., Green, A. J. \apj, 561, 264
\bibitem[Dickey \& Lockman (1990)]{dic90} Dickey, J. M., \& Lockman,
F. J. 1990, ARAA, 28, 215
\bibitem[Duncan \etal (1995)]{dun95} Duncan, A. R., Haynes, R. F.,
Stewart, R. T. \& Jones, K. L. 1995, MNRAS, 277, 319
\bibitem[English \etal (2000)]{eng00} English, J., Taylor, A. R.,
Mashchenko, S. Y., Irwin, J. A., Basu, S., \& Johnstone, Doug 2000,
\apjl, 533, L25
\bibitem[Finkbeiner (2003)]{fin03} Finkbeiner, D. 2003, \apjs, 146, 407
\bibitem[Forbes (2000)]{for00} Forbes, D. 2000, \apj, 120, 2594
\bibitem[Ghazzali, Joncas, \& Jean(1999)]{gha99} Ghazzali, 
N., Joncas, G., \& Jean, S.\ 1999, \apj, 511, 242
\bibitem[Gooch (1996)]{goo96} Gooch, R. E. 1996, Pub. Astron. Soc.
Aus., 14, 106
\bibitem[Goldman (2000)]{gol00}	Goldman, I.~2000, \apj, 541, 701
\bibitem[van Gorkom \& Ekers(1989)]{vang89} van Gorkom, J.~H.~\&
Ekers, R.~D.\ 1989, ASP Conf.~Ser.~6: Synthesis Imaging in Radio 
Astronomy, 341
\bibitem[Heiles, Reach \& Koo (1996)]{hei96} Heiles, C., Reach, W. T.,
Koo, B-C 1996, \apj, 466, 191
\bibitem[Heiles (1984)]{hei84} Heiles, C. 1984, \apjs, 55, 585
\bibitem[Heiles {\it et al.} (1980)]{hei80} Heiles, C., Chu, Y.-H., 
Troland, T.~H., Reynolds, R.~J., \& Yegingil, I.\ 1980, \apj, 242, 533
\bibitem[Heyer \etal~(1998)]{hey98}Heyer, M. H., Brunt, C. M.,
Snell, R. L., Howe, J. E., Schloerb, F. P., \& Carpenter, J. M.
1998, ApJS, 115, 241
\bibitem[Higgs \etal (2004)] {hig04} Higgs, L.A, Landecker, T.L.,
Asgekar, A.,  Davison, O.S., Rothwell, T.A., \& Yar-Uyaniker, A. 2004,
\aj, submitted.
\bibitem[Higgs \& Tapping (2000)] {hig00} Higgs, L. A, \& Tapping,
K. F. 2000, \aj, 120, 2471
\bibitem[Irwin \& Chaves (2003)]{irw03} Irwin, J. A. \& Chaves, T. 2003,
\apj, 585, 268
\bibitem[Kerton \& Martin (2000)]{ker00} Kerton, C. R., \& Martin, P. G.
2000, ApJS, 126, 85
\bibitem[Khalil, Joncas \& Nekka (2004)]{kha04} 
	Khalil, A., Joncas, G., \&  Nekka, F. 2004, \apj, 601, 352
\bibitem[Koo, Heiles \& Reach(1992)]{koo92} Koo, B-C, Heiles, C.,
\& Reach, W. T. 1992, \apj, 390, 108
\bibitem[Korpi \etal (1999)]{kor99} Korpi, M. J., Brandenburg, A.,
Shukurov, A., \& Tuominen, I. 1999, \aa, 350, 230
\bibitem[Landecker et al.(2000)]{lan00} Landecker, T.~L., Dewdney,
P. E., Burgess, T. A., Gray, A. D., Higgs, L. A., Hoffmann, A. P.,
Hovey, G. J., Karpa, D. R., Lacey, J. D., Prowse, N., Purton, C. R.,
Roger, R. S., Willis, A. G., Wyslouzil, W., Routledge, D.,
Vaneldik, J. F. 2000, \aaps, 145, 509
\bibitem[Landecker, Clutton-Brock \& Purton (1990)]{lan90}
Landecker, T. L., Clutton-Brock, M. \& Purton, C. R. 1990, \AA, 232, 207
\bibitem[Lockman, Hobbs, \& Shull (1986)]{loc86} Lockman, 
F.~J., Hobbs, L.~M., \& Shull, J.~M.\ 1986, \apj, 301, 380
\bibitem[Mac Low, McCray \& Norman (1989)]{mac89} Mac Low, M.-M.,
McCray, R., Norman, M. L. 1989, \apj, 337, 141
\bibitem[Matthews, Wallace \& Taylor (1998)]{mat98} Matthews, B. C.,
Wallace, B. J., \& Taylor, A. R. 1998, \apj, 493, 312
\bibitem[McClure-Griffiths \etal (2003)]{mcg03} McClure-Griffiths, N. M.,
Dickey, J. M., Gaensler, B. M., Green, A. J. 2003, ApJ, 594, 833
\bibitem[McClure-Griffiths \etal (2000)]{mcg00} McClure-Griffiths, N. M.,
Dickey, J. M., Gaensler, B. M., Green, A. J., Haynes, R. F., \&
Wieringa, M. H. 2000, ApJ, 119, 2828
\bibitem[McGlynn \etal (1996)]{mcg96}
McGlynn, T., Scollick, K., \& White, N.  1996,
      ``SkyView: The Multi-Wavelength Sky on the Internet",
      in New Horizons from Multi-Wavelength Sky Surveys, ed.s
      McLean, B.J. et al., Kluwer Academic Publishers, 
      IAU Symposium No. 179, p465.
\bibitem[M\"uller, Reif \& Reich (1987)]{mue87} M\"uller, P., Reif, K., \&
Reich, W. 1987, \aa, 183, 327
\bibitem[Norman \& Ikeuchi (1989)]{nor89} Norman, C. A., \& Ikeuchi, S.
1989, ApJ,  345, 372
\bibitem[Normandeau (2003)]{nor03} Normandeau, M., personal
communication
\bibitem[Normandeau, Taylor \& Dewdney (1996)]{nor96} Normandeau, M.,
Taylor, A. R., \& Dewdney, P. E. 1996, \nat, 380, 687
\bibitem[Normandeau, \etal (2000)]{nor00} Normandeau, M.,
Taylor, A. R., Dewdney, P. E., \& Basu, S. 2001, \apj, 119, 2982
\bibitem[Pineault \& Joncas (2000)]{pin00} Pineault S. \& Gilles, J. 
2000, 120, 3218
\bibitem[Reylolds (1989)]{rey89} Reynolds, R. J. 1989, \apj, 339, L29
\bibitem[Reylolds (1991)]{rey91} Reynolds, R. J. 1991, \apj, 372, L17
\bibitem[Routledge, Landecker \& Vaneldik (1986)]{rou86} Routledge,
D., Landecker, T. L. \& Vaneldik, J. F. 1986, MNRAS, 221, 809
\bibitem[Saken, Fesen \& Shull (1992)]{sak92} Saken, J. M., Fesen,
R. A., Shull, J. Michael 1992, \apjs, 81, 715
\bibitem[Sofui \& Reich (1979)]{sof79} Sofue, Y.~\& Reich, W.\ 1979,
\aaps, 38, 251 
\bibitem[Stil \& Irwin (2001)]{sti01} Stil, J. M., \& Irwin, J. A.
2001, \apj, 563, 816
\bibitem[Takahiro \& Basu (2003)]{tak03} Takahiro, J., \& Basu,
S. 2003, in preparation
\bibitem[Taylor \etal (2003)]{tay03} Taylor, A. R., Gibson, S. J.,
Peracaula, M., Martin, P. G., Landecker, T. L., Brunt, C. M., Dewdney,
P. E., Dougherty, S. M., Gray, A. D., Higgs, L. A., Kerton, C. R.,
Knee, L. B. G., Kothes, R., Purton, C. R., Uyaniker, B., Wallace,
B. J., Willis, A. G., \& Durand, D. 2003, \aj, 125, 3145
\bibitem[Tenorio-Tagle \etal (1987)]{ten87} Tenorio-Tagle, G., Franco,
J., Bodenheimer, P., \& Rozyczka, M. 1987, \aap, 179,219
\bibitem[Tomisaka (1990)]{tom90}Tomisaka, K. 1990, \apjl, 361, L5
\bibitem[Tomisaka (1998)]{tom98}Tomisaka, K. 1998, MNRAS, 298, 797
\bibitem[Uyaniker \& Kothes (2002)]{uya02} Uyaniker, B.~\& 
Kothes, R.\ 2002, \apj, 574, 805
\bibitem[Wolfire, McKee, Hollenbach, \& Tielens (2003)]{wol03}
Wolfire, M.~G., McKee, C.~F., Hollenbach, D., \& Tielens,
A.~G.~G.~M.\ 2003, \apj, 587, 278

\end{thebibliography}
\end{document}